\documentclass[%
 reprint,
 superscriptaddress,
nofootinbib,
 amsmath,amssymb,
 aps,
]{revtex4-2}

\usepackage{graphicx}
\usepackage{dcolumn}
\usepackage{bm}

\begin{document}


\title{Reevaluating reactor antineutrino spectra
with new measurements of the ratio between $^{235}$U and $^{239}$Pu $\beta$ spectra
}

\author{V.~Kopeikin}
\affiliation{National Research Centre Kurchatov Institute, 123182, Moscow, Russia}
\author{M.~Skorokhvatov}%
\affiliation{National Research Centre Kurchatov Institute, 123182, Moscow, Russia}
\affiliation{National Research Nuclear University MEPhI (Moscow Engineering Physics Institute), 115409, Moscow, Russia}
\author{O.~Titov}%
\email{titov\_oa@nrcki.ru}
\affiliation{National Research Centre Kurchatov Institute, 123182, Moscow, Russia}




\date{\today}

\begin{abstract}
We report a reanalysis of the reactor antineutrino energy spectra based on the new relative measurements of the ratio $R=\!^{e}S_5/^{e}S_9$ between cumulative $\beta$ spectra from $^{235}$U and $^{239}$Pu, performed at a research reactor in National Research Centre Kurchatov Institute (KI). A discrepancy with the $\beta$ spectra measured at Institut Laue-Langevin (ILL) was observed, indicating a steady excess of the ILL ratio by the factor of $1.054\pm0.002$. We find a value of the ratio between inverse beta decay cross section per fission for $^{235}$U and $^{239}$Pu: $(^5\sigma_f/^9\sigma_f)_{KI} = 1.45 \pm 0.03$, and then we reevaluate the converted antineutrino spectra for $^{235}$U and $^{238}$U. We conclude that the new predictions are consistent with the results of Daya Bay and STEREO experiments.
\end{abstract}

\maketitle



In reactor antineutrino studies, most experiments are analyzed on the basis of knowledge of the antineutrino spectra emitted from reactors. In these cases, the initial spectrum ''at the moment of birth'' in the reactor core is used. Uncertainties in the knowledge of this spectrum limit the sensitivity of the experiments, and systematic errors can simulate or mask new unexpected effects. In pressurized water reactors (PWR), the electron antineutrino ($\bar{\nu}_e$) flux emitted by nuclear fuels in the fission chain reaction is generated in $\beta$~decays of the neutron-rich fission fragments of U and Pu isotopes. Despite numerous past and current studies, the accurate specification of the energy $\bar{\nu}_e$ spectrum is an open problem.

The antineutrino spectra of $^{235}$U, $^{239}$Pu and $^{241}$Pu were predicted converting the cumulative $\beta$ spectra measured with the BILL spectrometer at the high flux reactor in Institut
Laue-Langevin (ILL)~\cite{Schreckenbach+_1981, von_Feilitzsch+_1982, Schreckenbach+_1985, Hahn+_1989}. The conversion procedure uses a set of about 30 allowed $\beta$ transitions to fit the measured electron spectrum.  The corresponding $\bar{\nu}_e$ spectra of these individual transitions are then summed to build the total $\bar{\nu}_e$ spectrum.  Later, a $\beta$-spectrum study for fast neutrons fissions of $^{238}$U was done in a separate experiment~\cite{Haag+_2014} at the neutron source FRM~II in Garching. In this case, a more empirical conversion method based on the similarity of the cumulative $\beta$ and $\bar{\nu}_e$ spectra was used.

There is another approach based on the summation method in which the antineutrino spectra from fission fragments are calculated \textit{ab initio} using $\beta$-decay information from nuclear databases and theoretical inputs. These calculations were used initially in several studies~\cite{King+_1958, *Borovoi+_1977, *Avignone+_1979} and are constantly being improved by means of new fission and nuclear data (see, e.~g., Ref.~\cite{Estienne+_2019} for recent results). Our study is focused on conversion calculations, so we do not consider the summation method hereinafter.

The commonly used conversion method is based on a scheme described in Refs.~\cite{Schreckenbach+_1981, von_Feilitzsch+_1982, Schreckenbach+_1985} and later in~\cite{Huber_2011, *Mueller+_2011}, where an improved model has been developed (Huber-Mueller model, HM). But the measured antineutrino rate in precise experiments near reactors via inverse beta decay (IBD) reaction
\[\bar{\nu}_e + p \to e^+ + n\]
(with a threshold of $1.8$~MeV) corresponds to a deficit of about $4-5\%$ as compared to the predicted rate. This anomaly, known as the Reactor Antineutrino Anomaly (RAA)~\cite{Mention+_2011}, caused a number of studies directed to searching $\bar{\nu}_e$ oscillations to a hypothetical, so-called ''sterile'', neutrino state. Alternatively, the explanation for RAA is due to possible errors in predicting the $\bar{\nu}_e$ spectrum for reactor antineutrino fluxes. 

A new analysis of the conversion procedure based on the recent measurements of the ratio between the cumulative fission $\beta$ spectra for $^{235}$U and $^{239}$Pu, performed in~\cite{Kopeikin+_2021}, is the topic of the present study.

Obviously, the HM model is strongly dependent on the original $\beta$ spectra measured in ILL experiments. These experiments consisted in irradiating thin targets of uranium and plutonium inserted to the reactor and exposed to a high neutron flux. The energies of $\beta$ particles were measured with a high precision by extracting electrons to the BILL magnetic spectrometer. There are two main specifications of the measured $\beta$ spectrum: the energy-dependent shape and the normalization connected with the number of $\beta$ particle per fission. The absolute rate and energy scale were obtained by means of calibrations, replacing fission targets with targets provided a known partial $(n, e^-)$ and $(n, \gamma)$ cross sections in order to measure internal-conversion electrons and $\beta$-decay electrons following neutron capture. 
	
For $^{235}$U, three measurements were performed at ILL under different reactor powers provided different signal-to-background conditions: 
\begin{itemize}
    \item 1981 – the first measurement at full reactor power of 57~MW; 
    \item 1982 – the test measurement at 8~MW without calibration;
    \item 1985 – the final measurement at 4~MW, but calibration has been performed at 57~MW.
\end{itemize}
All measurements show excellent agreement in the spectral shape, virtually coinciding within $1\%$. However, the absolute $\beta$ yields per fission of these spectra differed at the level of several percent, which has been associated with a number of reasons noted by the authors of Ref.~\cite{Schreckenbach+_1985}. The final measurement, generally used as a reference, exhibits the normalization uncertainty of $1.8\%$\footnote{Hereinafter, we present all uncertainties at~$1\sigma$ confidence level~($68\%$).}. For $^{239}$Pu, only one measurement of the cumulative $\beta$ spectrum was performed in 1982~\cite{von_Feilitzsch+_1982} at full reactor power of 57~MW. The calibration error according to the data~\cite{von_Feilitzsch+_1982} was $2.3\%$. The results of all the ILL $\beta$-spectra measurements were republished in 2014 with a finer grid~\cite{Haag+_2014_republication}. In the analysis below we use these data.

The study of long-term data in the Daya Bay reactor experiment, presented in \cite{DB_2017,DB_2019}, shows a deficit in the events rate caused by the yield of $^{235}$U antineutrinos, as compared with the HM prediction. At the same time, the rate associated with $^{239}$Pu antineutrinos is in a reasonable agreement. Similar results were obtained by the RENO collaboration~\cite{RENO_2019}. A lower antineutrino rate was also observed in the STEREO experiment~\cite{STEREO_2020} at the ILL reactor in an antineutrino flux from $^{235}$U.

Taking into account, as already mentioned, the good agreement of the shape of the measured $\beta$ spectra for $^{235}$U, it is most likely that systematical errors could have occurred during the absolute normalization procedure in $^{235}$U experiments, in particular, as a result of inconsistent calibration and measurements at different reactor powers, uncertainties in cross-section values for calibration targets, etc. The normalization was discussed, e.~g., in~\cite{Onillon+_2019}. While the preliminary results of Ref.~\cite{Onillon+_2019} show no clear evidence for the normalization bias, the authors point out that the data used for calibration (cross sections and internal conversion coefficients) underwent some changes since the ILL experiments. Therefore, it is important to reexamine the ILL results and to perform direct experimental measurements of the $\beta$-spectra ratio for $^{235}$U and $^{239}$Pu via a relative approach that is free of normalization uncertainties. As it has been shown in~\cite{Hayes+_2018}, this ratio
determines the relationship $(^5\sigma_f/^9\sigma_f)$ between the IBD cross sections integrated over the antineutrino spectra per fissions of $^{235}$U and $^{239}$Pu (for brevity, further on referred to as the ratio of IBD yields), that can be conveniently tested in reactor experiments.

Recently, preliminary results of an experiment at the research reactor IR-8 at NRC Kurchatov Institute (KI) were published~\cite{Kopeikin+_2021}. In this experiment, relative measurements of the ratio~$R$ between the cumulative fission $\beta$ spectra of $^{235}$U and $^{239}$Pu were carried out simultaneously, and therefore the neutron flux knowledge was not required to normalize the absolute rate of fission in the target. The ratio~$R$ was determined from the equation
\begin{equation} \label{Eq1}
R \equiv \frac{^{e}S_5}{^{e}S_9} = \frac{\sigma_9}{\sigma_5} \cdot \frac{N_9}{N_5} \cdot \frac{n_5}{n_9}, 
\end{equation}
where $i=5,\, 9$ refer to $^{235}$U and $^{239}$Pu, respectively. In Eq.~\eqref{Eq1}, $^{e}S_i$~is the cumulative $\beta$~spectrum, $\sigma_i$~is the fission cross section, $N_i$~is the number of nuclei in the target, $n_i$~is the detected spectrum of $\beta$~particles. 

The experiment~\cite{Kopeikin+_2021} uses foils of high-purity $^{235}$U and $^{239}$Pu. The fractions of $^{235}$U and $^{239}$Pu in the foils are known with a great precision; the error of the ratio $N_9 / N_5$ is~$0.2\%$.

The neutron flux in~\cite{Kopeikin+_2021} is thermal, with a small contribution of epithermal neutrons, so the error of the $\sigma_9 / \sigma_5$ ratio is mainly due to the cross section uncertainties for $^{235}$U and $^{239}$Pu fission by thermal neutrons. The deviation of the cross sections from the $1/v$ law and the correction related to a higher temperature of the neutron moderator ($42 ^{\circ}$~C) in the research reactor are also sources of uncertainty. The overall error of the $\sigma_9 / \sigma_5$~ratio is~$0.4\%$. Note that reasonable variation of the aforementioned corrections does not affect the value of~$\sigma_9 / \sigma_5$. The resulting systematic error of the ratio~\eqref{Eq1} is~$0.5\%$.

\begin{figure}[b]
\includegraphics[scale=0.5]{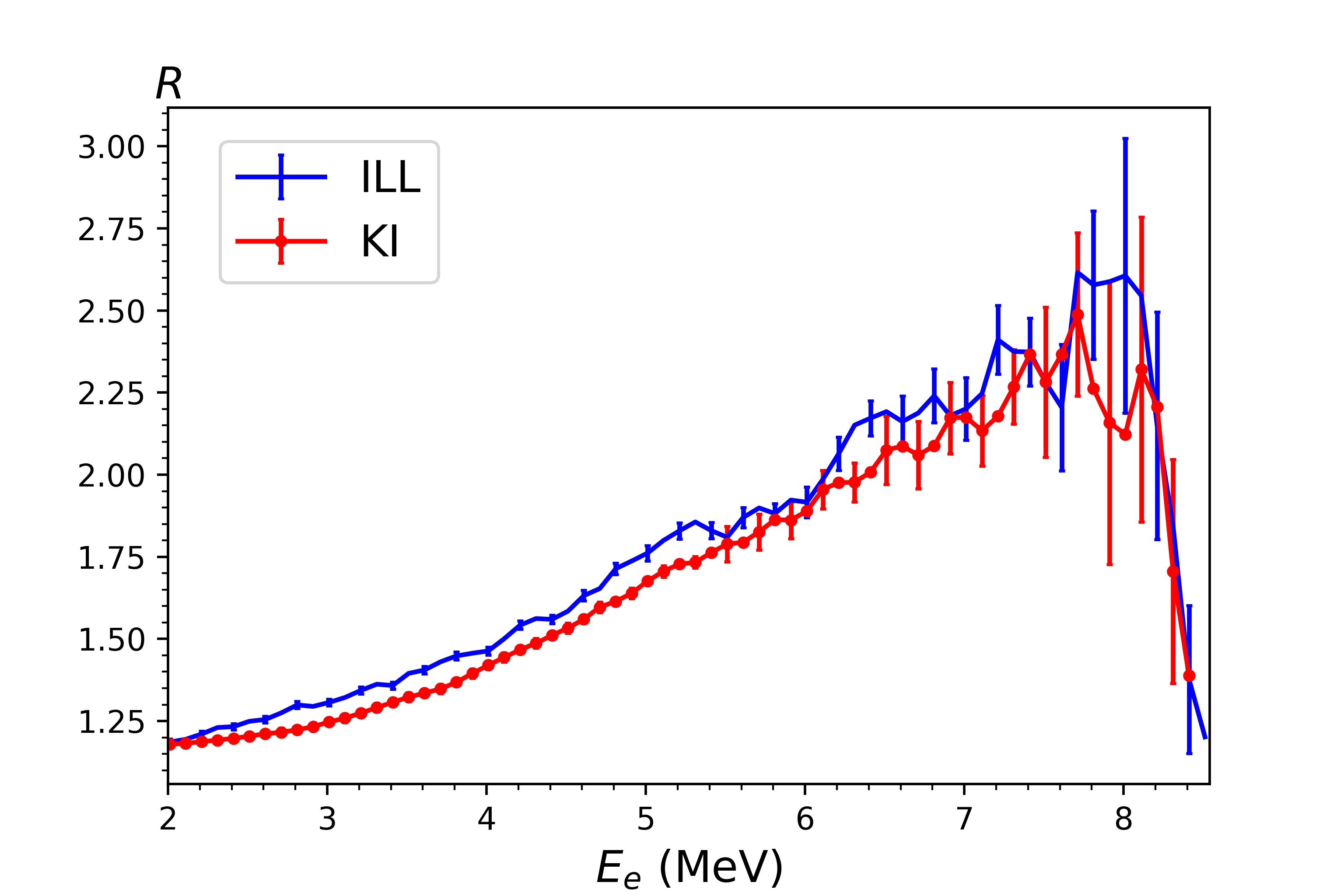}
\caption{\label{Fig_1} Ratios $R=\! ^{e}S_5/^{e}S_9$ between cumulative $\beta$ spectra from $^{235}$U and $^{239}$Pu from ILL data~\cite{Haag+_2014_republication} (the upper curve, blue) and KI data~\cite{Kopeikin+_2021} (the lower curve, red). Total electron energies are given. Only statistical errors are shown.}
\end{figure}

A more detailed description of the experiment and its results can be found in Ref.~\cite{Kopeikin+_2021}. Here in Fig.~\ref{Fig_1}, we present the results of the KI experiment in comparison with the values calculated from the ILL data for $^{235}$U and $^{239}$Pu republished in~\cite{Haag+_2014_republication}.

\begin{figure}
\includegraphics[scale=0.5]{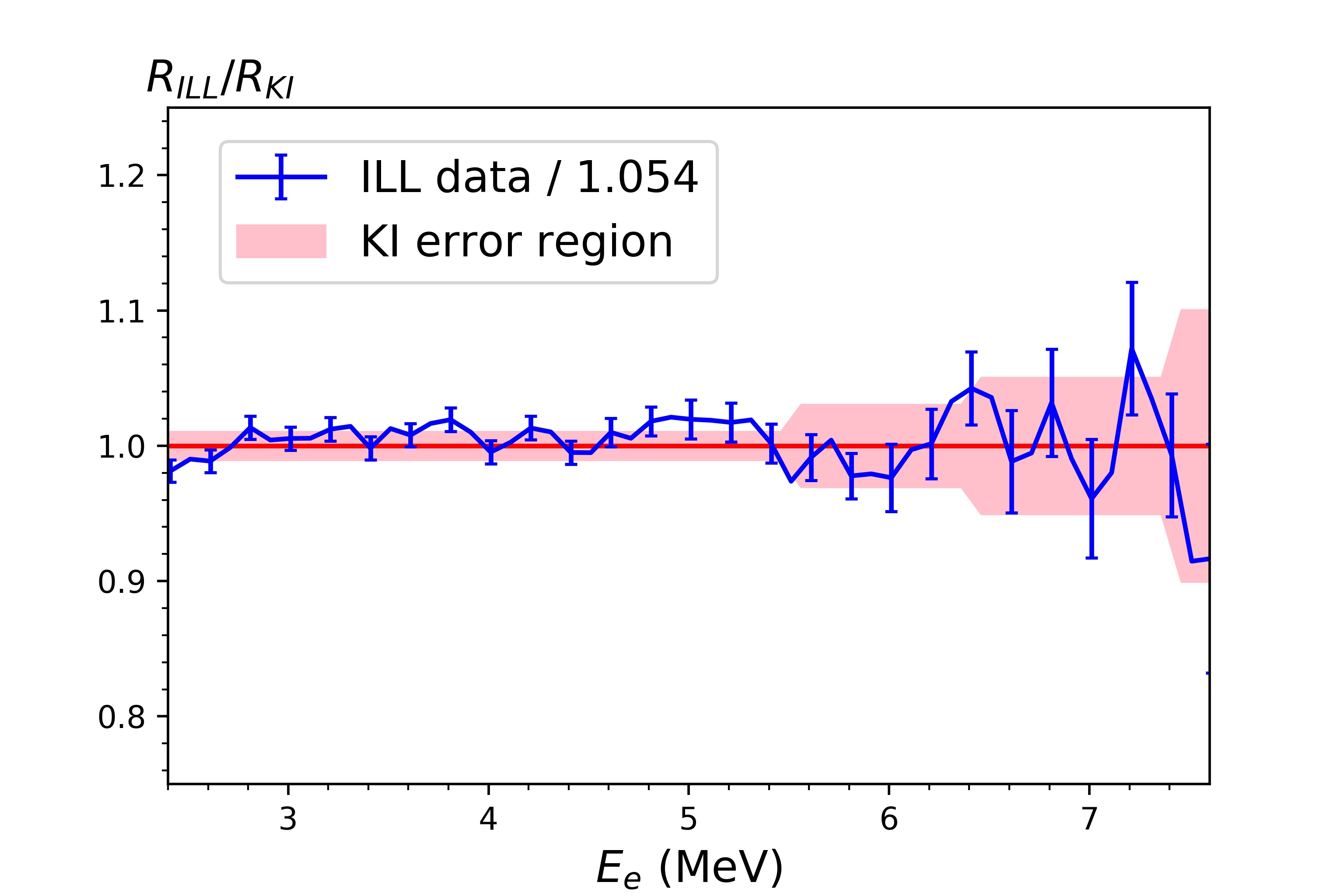}
\caption{\label{Fig_2} Ratios $R$ between cumulative $\beta$ spectra from $^{235}$U and $^{239}$Pu, normalized to the KI data. Plotted ILL quantities were divided by $1.054$, as explained in the text. The colored region shows KI uncertainties.}
\end{figure}

In Fig.~\ref{Fig_1}, the horizontal axis corresponds to the total electron energy $E_e$ (including the electron mass) and only statistical errors for the data are presented. One can clearly see that the KI measurements give a lower value for the ratio. Nevertheless, the shapes of the curves are consistent with each other and the difference between them in the region of $2.5-7.5$~MeV remains practically constant. At higher energies, the statistical uncertainty becomes too large, but a simultaneous sharp drop in values under the same energies within the region of $8-8.5$~MeV indicates a good correspondence of the energy scales between KI and ILL measurements. A steady excess of the ILL ratio is a strong argument for a revision of the currently used  $\bar{\nu}_e$-spectra reconstruction, mainly for $^{235}$U isotope. Using the weighted least squares procedure, one can fit this excess by a constant factor, which was determined to be $k = 1.054 \pm 0.002$. For more visibility, all relative data normalized to KI ratio were presented in Fig.~\ref{Fig_2}, where plotted ILL quantities were divided by $k$ and superposed on the KI data taken with a highlighted uncertainty region. As one can see from the figure, the obtained values agree in the full energy range. 

Analysis and comparison of the new experimental data allow us to derive the following consequences.
\begin{description}
\item[A] 
As far as the ratio $(^5\sigma_f / ^9\sigma_f)$ between the IBD cross sections per the $^{235}$U and $^{239}$Pu fissions is strongly constrained by the ratio of the $^{235}$U to $^{239}$Pu $\beta$ spectra~\cite{Hayes+_2018}, we find that $(^5\sigma_f/^9\sigma_f)_{KI}$ is about $5.4\%$ lower than the HM result. The obtained value $(^5\sigma_f/^9\sigma_f)_{KI} = 1.45 \pm 0.03$ is in a good agreement with the combined Daya Bay and RENO result $(^5\sigma_f/^9\sigma_f) =  1.44 \pm 0.10$~\cite{DB_2019, RENO_2019}.
\item [B] 
Based on Daya Bay and STEREO results noted above, we believe that the difference between ILL and KI ratios is due to an overestimation in the absolute normalization of the ILL $\beta$ spectrum for $^{235}$U. Since the antineutrino spectrum for $^{235}$U was derived converting the $\beta$ spectrum, it has to be corrected in the same way. In Table~\ref{Table1}, a new version of the antineutrino spectrum for $^{235}$U obtained by means of the HM conversion method is presented. 
\item [C] 
The cumulative $\beta$ spectrum for $^{238}$U~\cite{Haag+_2014} was obtained using experimental data that were normalized to the BILL measurement~\cite{Schreckenbach+_1985} for $^{235}$U. Since this normalization leads to a correlation between the $^{238}$U spectrum obtained in the experiment with the spectrum of $^{235}$U~\cite{Haag+_2014}, we have to renormalize the $\beta$ spectrum of $^{238}$U in accordance with the correction described above. The corresponding antineutrino spectrum of $^{238}$U is shown in Table~\ref{Table1}. For energies below 3 MeV, the spectrum was continued according to Ref.~\cite{Kopeikin_2012}.
\end{description}

\begin{table}
\caption{\label{Table1}%
Reevaluated antineutrino spectra $N_{\bar{\nu}_e}$ \mbox{(MeV$^{-1} \cdot$ fission$^{-1}$)} for $^{235}$U and $^{238}$U. For $^{235}$U, the errors~$\delta$ are taken from~\cite{Huber_2011} and include statistical and conversion uncertainties. The errors~$\delta$ of the $^{238}$U spectrum include statistical, conversion and normalization uncertainties of the experiment~\cite{Haag+_2014}. The absolute normalization error ($1.8\%$) of the ILL experiment~\cite{Schreckenbach+_1985} is not included.}
\begin{ruledtabular}
\begin{tabular}{ccccc}
$E\,$\text{(MeV)} &\multicolumn{2}{c}{$^{235}$\text{U}} & \multicolumn{2}{c}{$^{238}$\text{U}}\\
& $N_{\bar{\nu}_e}$ & $\delta$ (\%) & $N_{\bar{\nu}_e}$ & $\delta$ (\%)\\
\hline
2.0 & 1.25 & $<1.0$ & 1.54 & $\sim 4.5$\\
2.25&	1.06 &	 &	1.35 & \\	
2.5 &	8.68$\times 10^{-1}$&	 &	1.18 & \\	
2.75&	7.31$\times 10^{-1}$&	 &	1.04 & \\
3.0	&   6.18$\times 10^{-1}$&	&	9.10$\times 10^{-1}$&	 \\
3.25&	5.25$\times 10^{-1}$&	&	7.55$\times 10^{-1}$& 4.2 \\
3.5	&   4.31$\times 10^{-1}$&	&   6.27$\times 10^{-1}$ &	3.9 \\
3.75&	3.45$\times 10^{-1}$&	1.1	&   5.13$\times 10^{-1}$&	3.9\\
4.0&	2.79$\times 10^{-1}$&	1.2	&   4.21$\times 10^{-1}$	&3.9\\
4.25&	2.18$\times 10^{-1}$&	1.4&	3.32$\times 10^{-1}$&	4.0\\
4.5&	1.70$\times 10^{-1}$&	1.7&	2.64$\times 10^{-1}$&	4.1\\
4.75&	1.31$\times 10^{-1}$&	1.8&	2.06$\times 10^{-1}$&	4.4\\
5.0&	1.04$\times 10^{-1}$&	1.9&    1.61$\times 10^{-1}$&	4.7\\
5.25&	8.20$\times 10^{-2}$&	2.0&	1.27$\times 10^{-1}$&	5.0\\
5.5&	6.13$\times 10^{-2}$&	2.2&	9.79$\times 10^{-2}$&	5.8\\
5.75&	4.84$\times 10^{-2}$&	2.4&	7.34$\times 10^{-2}$&	6.6\\
6.0&	3.69$\times 10^{-2}$&	2.7&	5.33$\times 10^{-2}$&	8.1\\
6.25&	2.72$\times 10^{-2}$&	3.0&     3.77$\times 10^{-2}$&	11\\
6.5&	2.06$\times 10^{-2}$&	3.0&	2.89$\times 10^{-2}$&	13\\
6.75&	1.53$\times 10^{-2}$&	3.3&	2.66$\times 10^{-2}$&	12\\
7.0&	1.08$\times 10^{-2}$&	3.6&	1.99$\times 10^{-2}$&	14\\
7.25&	6.80$\times 10^{-3}$&	4.1&	1.08$\times 10^{-2}$&	22\\
7.5&	4.40$\times 10^{-3}$&	4.4&	6.77$\times 10^{-3}$&	30\\
7.75&	2.82$\times 10^{-3}$&	5.0&	4.70$\times 10^{-3}$& $\sim 30$\\
8.0&	1.54$\times 10^{-3}$&	7.0&	3.0$\times 10^{-3}$&
\end{tabular}
\end{ruledtabular}
\end{table}

Finally, we reevaluate the expected IBD yields for reactor experiments. Taking the corrected spectra for $^{235}$U and $^{238}$U and the unmodified HM spectra for $^{239}$Pu and $^{241}$Pu, we obtain the following IBD yields for each component: 
$^{235}\sigma_f= (6.27 \pm 0.13)\times 10^{-43}\, \text{cm$^2$/fission},$
$^{239}\sigma_f= (4.33 \pm 0.11)\times 10^{-43}\, \text{cm$^2$/fission},$
$^{238}\sigma_f= (9.34 \pm 0.47)\times 10^{-43}\, \text{cm$^2$/fission},$
$^{241}\sigma_f= (6.01 \pm 0.13)\times 10^{-43}\, \text{cm$^2$/fission}.$
The IBD cross sections were calculated according to Ref.~\cite{Fayans_1985, *Vogel_1984}. With the information provided in Refs.~\cite{DB_2019} and~\cite{STEREO_2020}, we have reanalyzed the ratios between the measured and the predicted reactor antineutrino yields for an average composition of the reactor fuel. The Daya Bay and STEREO IBD yields \mbox{$\langle \sigma_f \rangle = (5.94 \pm 0.09) \times 10^{-43}$ cm$^2$/fission} and \mbox{$^{235}\sigma_f= (6.34 \pm 0.22)\times 10^{-43}$ cm$^2$/fission} correspond respectively to $(1.007 \pm 0.031)$ and $(1.012 \pm 0.043)$
of our predictions.

The derivation of the total antineutrino flux considered here did not take into a few small corrections, associated with the ''off-equilibrium'' effect, neutron captures by the fission products and the antineutrino radiation from the stored spent reactor fuel. Estimated values of these corrections can be found in~\cite{Kopeikin+_2017}. A thorough analysis of the reactor antineutrinos, based on new high statistic experimental data for $\beta$ spectra of the contributing fission isotopes, is in progress now.

\begin{acknowledgments}
We are grateful to L.~Oberauer and N.~Haag for attention to this work and useful comments.
\end{acknowledgments}

\bibliography{references}

\providecommand{\noopsort}[1]{}\providecommand{\singleletter}[1]{#1}%
\begin{thebibliography}{24}%
\makeatletter
\providecommand \@ifxundefined [1]{%
 \@ifx{#1\undefined}
}%
\providecommand \@ifnum [1]{%
 \ifnum #1\expandafter \@firstoftwo
 \else \expandafter \@secondoftwo
 \fi
}%
\providecommand \@ifx [1]{%
 \ifx #1\expandafter \@firstoftwo
 \else \expandafter \@secondoftwo
 \fi
}%
\providecommand \natexlab [1]{#1}%
\providecommand \enquote  [1]{``#1''}%
\providecommand \bibnamefont  [1]{#1}%
\providecommand \bibfnamefont [1]{#1}%
\providecommand \citenamefont [1]{#1}%
\providecommand \href@noop [0]{\@secondoftwo}%
\providecommand \href [0]{\begingroup \@sanitize@url \@href}%
\providecommand \@href[1]{\@@startlink{#1}\@@href}%
\providecommand \@@href[1]{\endgroup#1\@@endlink}%
\providecommand \@sanitize@url [0]{\catcode `\\12\catcode `\$12\catcode
  `\&12\catcode `\#12\catcode `\^12\catcode `\_12\catcode `\%12\relax}%
\providecommand \@@startlink[1]{}%
\providecommand \@@endlink[0]{}%
\providecommand \url  [0]{\begingroup\@sanitize@url \@url }%
\providecommand \@url [1]{\endgroup\@href {#1}{\urlprefix }}%
\providecommand \urlprefix  [0]{URL }%
\providecommand \Eprint [0]{\href }%
\providecommand \doibase [0]{https://doi.org/}%
\providecommand \selectlanguage [0]{\@gobble}%
\providecommand \bibinfo  [0]{\@secondoftwo}%
\providecommand \bibfield  [0]{\@secondoftwo}%
\providecommand \translation [1]{[#1]}%
\providecommand \BibitemOpen [0]{}%
\providecommand \bibitemStop [0]{}%
\providecommand \bibitemNoStop [0]{.\EOS\space}%
\providecommand \EOS [0]{\spacefactor3000\relax}%
\providecommand \BibitemShut  [1]{\csname bibitem#1\endcsname}%
\let\auto@bib@innerbib\@empty
\bibitem [{\citenamefont {Schreckenbach}\ \emph {et~al.}(1981)\citenamefont
  {Schreckenbach}, \citenamefont {Faust}, \citenamefont {{von Feilitzsch}},
  \citenamefont {Hahn}, \citenamefont {Hawerkamp},\ and\ \citenamefont
  {Vuilleumier}}]{Schreckenbach+_1981}%
  \BibitemOpen
  \bibfield  {author} {\bibinfo {author} {\bibfnamefont {K.}~\bibnamefont
  {Schreckenbach}}, \bibinfo {author} {\bibfnamefont {H.}~\bibnamefont
  {Faust}}, \bibinfo {author} {\bibfnamefont {F.}~\bibnamefont {{von
  Feilitzsch}}}, \bibinfo {author} {\bibfnamefont {A.}~\bibnamefont {Hahn}},
  \bibinfo {author} {\bibfnamefont {K.}~\bibnamefont {Hawerkamp}},\ and\
  \bibinfo {author} {\bibfnamefont {J.}~\bibnamefont {Vuilleumier}},\
  }\bibfield  {title} {\bibinfo {title} {Absolute measurement of the beta
  spectrum from $^{235}${U} fission as a basis for reactor antineutrino
  experiments},\ }\href
  {https://doi.org/https://doi.org/10.1016/0370-2693(81)91120-5} {\bibfield
  {journal} {\bibinfo  {journal} {Phys. Lett. B}\ }\textbf {\bibinfo {volume}
  {99}},\ \bibinfo {pages} {251} (\bibinfo {year} {1981})}\BibitemShut
  {NoStop}%
\bibitem [{\citenamefont {{von Feilitzsch}}\ \emph {et~al.}(1982)\citenamefont
  {{von Feilitzsch}}, \citenamefont {Hahn},\ and\ \citenamefont
  {Schreckenbach}}]{von_Feilitzsch+_1982}%
  \BibitemOpen
  \bibfield  {author} {\bibinfo {author} {\bibfnamefont {F.}~\bibnamefont {{von
  Feilitzsch}}}, \bibinfo {author} {\bibfnamefont {A.}~\bibnamefont {Hahn}},\
  and\ \bibinfo {author} {\bibfnamefont {K.}~\bibnamefont {Schreckenbach}},\
  }\bibfield  {title} {\bibinfo {title} {Experimental beta-spectra from
  $^{239}${P}u and $^{235}${U} thermal neutron fission products and their
  correlated antineutrino spectra},\ }\href
  {https://doi.org/https://doi.org/10.1016/0370-2693(82)90622-0} {\bibfield
  {journal} {\bibinfo  {journal} {Phys. Lett. B}\ }\textbf {\bibinfo {volume}
  {118}},\ \bibinfo {pages} {162} (\bibinfo {year} {1982})}\BibitemShut
  {NoStop}%
\bibitem [{\citenamefont {Schreckenbach}\ \emph {et~al.}(1985)\citenamefont
  {Schreckenbach}, \citenamefont {Colvin}, \citenamefont {Gelletly},\ and\
  \citenamefont {{Von Feilitzsch}}}]{Schreckenbach+_1985}%
  \BibitemOpen
  \bibfield  {author} {\bibinfo {author} {\bibfnamefont {K.}~\bibnamefont
  {Schreckenbach}}, \bibinfo {author} {\bibfnamefont {G.}~\bibnamefont
  {Colvin}}, \bibinfo {author} {\bibfnamefont {W.}~\bibnamefont {Gelletly}},\
  and\ \bibinfo {author} {\bibfnamefont {F.}~\bibnamefont {{Von Feilitzsch}}},\
  }\bibfield  {title} {\bibinfo {title} {Determination of the antineutrino
  spectrum from $^{235}${U} thermal neutron fission products up
  to~9.5~{M}e{V}},\ }\href
  {https://doi.org/https://doi.org/10.1016/0370-2693(85)91337-1} {\bibfield
  {journal} {\bibinfo  {journal} {Phys. Lett. B}\ }\textbf {\bibinfo {volume}
  {160}},\ \bibinfo {pages} {325} (\bibinfo {year} {1985})}\BibitemShut
  {NoStop}%
\bibitem [{\citenamefont {Hahn}\ \emph {et~al.}(1989)\citenamefont {Hahn},
  \citenamefont {Schreckenbach}, \citenamefont {Gelletly}, \citenamefont {{von
  Feilitzsch}}, \citenamefont {Colvin},\ and\ \citenamefont
  {Krusche}}]{Hahn+_1989}%
  \BibitemOpen
  \bibfield  {author} {\bibinfo {author} {\bibfnamefont {A.}~\bibnamefont
  {Hahn}}, \bibinfo {author} {\bibfnamefont {K.}~\bibnamefont {Schreckenbach}},
  \bibinfo {author} {\bibfnamefont {W.}~\bibnamefont {Gelletly}}, \bibinfo
  {author} {\bibfnamefont {F.}~\bibnamefont {{von Feilitzsch}}}, \bibinfo
  {author} {\bibfnamefont {G.}~\bibnamefont {Colvin}},\ and\ \bibinfo {author}
  {\bibfnamefont {B.}~\bibnamefont {Krusche}},\ }\bibfield  {title} {\bibinfo
  {title} {Antineutrino spectra from $^{241}${P}u and $^{239}${P}u thermal
  neutron fission products},\ }\href
  {https://doi.org/https://doi.org/10.1016/0370-2693(89)91598-0} {\bibfield
  {journal} {\bibinfo  {journal} {Phys. Lett. B}\ }\textbf {\bibinfo {volume}
  {218}},\ \bibinfo {pages} {365} (\bibinfo {year} {1989})}\BibitemShut
  {NoStop}%
\bibitem [{\citenamefont {Haag}\ \emph
  {et~al.}(2014{\natexlab{a}})\citenamefont {Haag}, \citenamefont {G\"utlein},
  \citenamefont {Hofmann}, \citenamefont {Oberauer}, \citenamefont {Potzel},
  \citenamefont {Schreckenbach},\ and\ \citenamefont {Wagner}}]{Haag+_2014}%
  \BibitemOpen
  \bibfield  {author} {\bibinfo {author} {\bibfnamefont {N.}~\bibnamefont
  {Haag}}, \bibinfo {author} {\bibfnamefont {A.}~\bibnamefont {G\"utlein}},
  \bibinfo {author} {\bibfnamefont {M.}~\bibnamefont {Hofmann}}, \bibinfo
  {author} {\bibfnamefont {L.}~\bibnamefont {Oberauer}}, \bibinfo {author}
  {\bibfnamefont {W.}~\bibnamefont {Potzel}}, \bibinfo {author} {\bibfnamefont
  {K.}~\bibnamefont {Schreckenbach}},\ and\ \bibinfo {author} {\bibfnamefont
  {F.~M.}\ \bibnamefont {Wagner}},\ }\bibfield  {title} {\bibinfo {title}
  {Experimental determination of the antineutrino spectrum of the fission
  products of $^{238}\mathrm{U}$},\ }\href
  {https://doi.org/10.1103/PhysRevLett.112.122501} {\bibfield  {journal}
  {\bibinfo  {journal} {Phys. Rev. Lett.}\ }\textbf {\bibinfo {volume} {112}},\
  \bibinfo {pages} {122501} (\bibinfo {year} {2014}{\natexlab{a}})},\ \Eprint
  {https://arxiv.org/abs/1312.5601} {arXiv:1312.5601 [nucl-ex]} \BibitemShut
  {NoStop}%
\bibitem [{\citenamefont {King}\ and\ \citenamefont
  {Perkins}(1958)}]{King+_1958}%
  \BibitemOpen
  \bibfield  {author} {\bibinfo {author} {\bibfnamefont {R.~W.}\ \bibnamefont
  {King}}\ and\ \bibinfo {author} {\bibfnamefont {J.~F.}\ \bibnamefont
  {Perkins}},\ }\bibfield  {title} {\bibinfo {title} {Inverse beta decay and
  the two-component neutrino},\ }\href
  {https://doi.org/10.1103/PhysRev.112.963} {\bibfield  {journal} {\bibinfo
  {journal} {Phys. Rev.}\ }\textbf {\bibinfo {volume} {112}},\ \bibinfo {pages}
  {963} (\bibinfo {year} {1958})}\BibitemShut {NoStop}%
\bibitem [{\citenamefont {Borovoi}\ \emph {et~al.}(1977)\citenamefont
  {Borovoi}, \citenamefont {Dobrynin},\ and\ \citenamefont
  {Kopeikin}}]{Borovoi+_1977}%
  \BibitemOpen
  \bibfield  {author} {\bibinfo {author} {\bibfnamefont {A.~A.}\ \bibnamefont
  {Borovoi}}, \bibinfo {author} {\bibfnamefont {Y.~L.}\ \bibnamefont
  {Dobrynin}},\ and\ \bibinfo {author} {\bibfnamefont {V.~I.}\ \bibnamefont
  {Kopeikin}},\ }\bibfield  {title} {\bibinfo {title} {Energy spectra of
  electrons and antineutrinos from fission fragments of $^{235}${U} and
  $^{239}${P}u},\ }\href@noop {} {\bibfield  {journal} {\bibinfo  {journal}
  {Sov. J. Nucl. Phys.}\ }\textbf {\bibinfo {volume} {25}},\ \bibinfo {pages}
  {144} (\bibinfo {year} {1977})}\BibitemShut {NoStop}%
\bibitem [{\citenamefont {Avignone}\ \emph {et~al.}(1979)\citenamefont
  {Avignone}, \citenamefont {Hopkins},\ and\ \citenamefont
  {Greenwood}}]{Avignone+_1979}%
  \BibitemOpen
  \bibfield  {author} {\bibinfo {author} {\bibfnamefont {F.~T.~I.}\
  \bibnamefont {Avignone}}, \bibinfo {author} {\bibfnamefont {L.~P.}\
  \bibnamefont {Hopkins}},\ and\ \bibinfo {author} {\bibfnamefont {Z.~D.}\
  \bibnamefont {Greenwood}},\ }\bibfield  {title} {\bibinfo {title}
  {Theoretical beta spectrum from uranium-235 fission fragments in secular
  equilibrium},\ }\href {https://doi.org/10.13182/NSE79-A19465} {\bibfield
  {journal} {\bibinfo  {journal} {Nucl. Sci. Eng.}\ }\textbf {\bibinfo {volume}
  {72}},\ \bibinfo {pages} {216} (\bibinfo {year} {1979})}\BibitemShut
  {NoStop}%
\bibitem [{\citenamefont {Estienne}\ \emph {et~al.}(2019)\citenamefont
  {Estienne} \emph {et~al.}}]{Estienne+_2019}%
  \BibitemOpen
  \bibfield  {author} {\bibinfo {author} {\bibfnamefont {M.}~\bibnamefont
  {Estienne}} \emph {et~al.},\ }\bibfield  {title} {\bibinfo {title} {Updated
  summation model: An improved agreement with the daya bay antineutrino
  fluxes},\ }\href {https://doi.org/10.1103/PhysRevLett.123.022502} {\bibfield
  {journal} {\bibinfo  {journal} {Phys. Rev. Lett.}\ }\textbf {\bibinfo
  {volume} {123}},\ \bibinfo {pages} {022502} (\bibinfo {year}
  {2019})}\BibitemShut {NoStop}%
\bibitem [{\citenamefont {Huber}(2011)}]{Huber_2011}%
  \BibitemOpen
  \bibfield  {author} {\bibinfo {author} {\bibfnamefont {P.}~\bibnamefont
  {Huber}},\ }\bibfield  {title} {\bibinfo {title} {Determination of
  antineutrino spectra from nuclear reactors},\ }\href
  {https://doi.org/10.1103/PhysRevC.84.024617} {\bibfield  {journal} {\bibinfo
  {journal} {Phys. Rev. C}\ }\textbf {\bibinfo {volume} {84}},\ \bibinfo
  {pages} {024617} (\bibinfo {year} {2011})}\BibitemShut {NoStop}%
\bibitem [{\citenamefont {Mueller}\ \emph {et~al.}(2011)\citenamefont
  {Mueller}, \citenamefont {Lhuillier}, \citenamefont {Fallot}, \citenamefont
  {Letourneau}, \citenamefont {Cormon}, \citenamefont {Fechner}, \citenamefont
  {Giot}, \citenamefont {Lasserre}, \citenamefont {Martino}, \citenamefont
  {Mention}, \citenamefont {Porta},\ and\ \citenamefont
  {Yermia}}]{Mueller+_2011}%
  \BibitemOpen
  \bibfield  {author} {\bibinfo {author} {\bibfnamefont {T.~A.}\ \bibnamefont
  {Mueller}}, \bibinfo {author} {\bibfnamefont {D.}~\bibnamefont {Lhuillier}},
  \bibinfo {author} {\bibfnamefont {M.}~\bibnamefont {Fallot}}, \bibinfo
  {author} {\bibfnamefont {A.}~\bibnamefont {Letourneau}}, \bibinfo {author}
  {\bibfnamefont {S.}~\bibnamefont {Cormon}}, \bibinfo {author} {\bibfnamefont
  {M.}~\bibnamefont {Fechner}}, \bibinfo {author} {\bibfnamefont
  {L.}~\bibnamefont {Giot}}, \bibinfo {author} {\bibfnamefont {T.}~\bibnamefont
  {Lasserre}}, \bibinfo {author} {\bibfnamefont {J.}~\bibnamefont {Martino}},
  \bibinfo {author} {\bibfnamefont {G.}~\bibnamefont {Mention}}, \bibinfo
  {author} {\bibfnamefont {A.}~\bibnamefont {Porta}},\ and\ \bibinfo {author}
  {\bibfnamefont {F.}~\bibnamefont {Yermia}},\ }\bibfield  {title} {\bibinfo
  {title} {Improved predictions of reactor antineutrino spectra},\ }\href
  {https://doi.org/10.1103/PhysRevC.83.054615} {\bibfield  {journal} {\bibinfo
  {journal} {Phys. Rev. C}\ }\textbf {\bibinfo {volume} {83}},\ \bibinfo
  {pages} {054615} (\bibinfo {year} {2011})}\BibitemShut {NoStop}%
\bibitem [{\citenamefont {Mention}\ \emph {et~al.}(2011)\citenamefont
  {Mention}, \citenamefont {Fechner}, \citenamefont {Lasserre}, \citenamefont
  {Mueller}, \citenamefont {Lhuillier}, \citenamefont {Cribier},\ and\
  \citenamefont {Letourneau}}]{Mention+_2011}%
  \BibitemOpen
  \bibfield  {author} {\bibinfo {author} {\bibfnamefont {G.}~\bibnamefont
  {Mention}}, \bibinfo {author} {\bibfnamefont {M.}~\bibnamefont {Fechner}},
  \bibinfo {author} {\bibfnamefont {T.}~\bibnamefont {Lasserre}}, \bibinfo
  {author} {\bibfnamefont {T.~A.}\ \bibnamefont {Mueller}}, \bibinfo {author}
  {\bibfnamefont {D.}~\bibnamefont {Lhuillier}}, \bibinfo {author}
  {\bibfnamefont {M.}~\bibnamefont {Cribier}},\ and\ \bibinfo {author}
  {\bibfnamefont {A.}~\bibnamefont {Letourneau}},\ }\bibfield  {title}
  {\bibinfo {title} {Reactor antineutrino anomaly},\ }\href
  {https://doi.org/10.1103/PhysRevD.83.073006} {\bibfield  {journal} {\bibinfo
  {journal} {Phys. Rev. D}\ }\textbf {\bibinfo {volume} {83}},\ \bibinfo
  {pages} {073006} (\bibinfo {year} {2011})}\BibitemShut {NoStop}%
\bibitem [{\citenamefont {Kopeikin}\ \emph {et~al.}(2021)\citenamefont
  {Kopeikin}, \citenamefont {Panin},\ and\ \citenamefont
  {Sabelnikov}}]{Kopeikin+_2021}%
  \BibitemOpen
  \bibfield  {author} {\bibinfo {author} {\bibfnamefont {V.~I.}\ \bibnamefont
  {Kopeikin}}, \bibinfo {author} {\bibfnamefont {Y.~N.}\ \bibnamefont
  {Panin}},\ and\ \bibinfo {author} {\bibfnamefont {A.~A.}\ \bibnamefont
  {Sabelnikov}},\ }\bibfield  {title} {\bibinfo {title} {Measurement of the
  ratio of cumulative spectra of beta particles from $^{235}${U} and
  $^{239}${P}u fission products for solving problems of reactor-antineutrino
  physics},\ }\href@noop {} {\bibfield  {journal} {\bibinfo  {journal} {Physics
  of Atomic Nuclei}\ }\textbf {\bibinfo {volume} {84}},\ \bibinfo {pages} {1}
  (\bibinfo {year} {2021})}\BibitemShut {NoStop}%
\bibitem [{\citenamefont {Haag}\ \emph
  {et~al.}(2014{\natexlab{b}})\citenamefont {Haag}, \citenamefont {Gelletly},
  \citenamefont {von Feilitzsch}, \citenamefont {Oberauer}, \citenamefont
  {Potzel}, \citenamefont {Schreckenbach},\ and\ \citenamefont
  {Sonzogni}}]{Haag+_2014_republication}%
  \BibitemOpen
  \bibfield  {author} {\bibinfo {author} {\bibfnamefont {N.}~\bibnamefont
  {Haag}}, \bibinfo {author} {\bibfnamefont {W.}~\bibnamefont {Gelletly}},
  \bibinfo {author} {\bibfnamefont {F.}~\bibnamefont {von Feilitzsch}},
  \bibinfo {author} {\bibfnamefont {L.}~\bibnamefont {Oberauer}}, \bibinfo
  {author} {\bibfnamefont {W.}~\bibnamefont {Potzel}}, \bibinfo {author}
  {\bibfnamefont {K.}~\bibnamefont {Schreckenbach}},\ and\ \bibinfo {author}
  {\bibfnamefont {A.~A.}\ \bibnamefont {Sonzogni}},\ }\href@noop {} {\bibinfo
  {title} {Re-publication of the data from the {BILL} magnetic spectrometer:
  The cumulative $\beta$ spectra of the fission products of $^{235}${U},
  $^{239}${P}u, and $^{241}${P}u}} (\bibinfo {year} {2014}{\natexlab{b}}),\
  \Eprint {https://arxiv.org/abs/1405.3501} {arXiv:1405.3501 [nucl-ex]}
  \BibitemShut {NoStop}%
\bibitem [{\citenamefont {An}\ \emph {et~al.}(2017)\citenamefont {An} \emph
  {et~al.}}]{DB_2017}%
  \BibitemOpen
  \bibfield  {author} {\bibinfo {author} {\bibfnamefont {F.~P.}\ \bibnamefont
  {An}} \emph {et~al.} (\bibinfo {collaboration} {Daya Bay Collaboration}),\
  }\bibfield  {title} {\bibinfo {title} {Evolution of the reactor antineutrino
  flux and spectrum at {D}aya {B}ay},\ }\href
  {https://doi.org/10.1103/PhysRevLett.118.251801} {\bibfield  {journal}
  {\bibinfo  {journal} {Phys. Rev. Lett.}\ }\textbf {\bibinfo {volume} {118}},\
  \bibinfo {pages} {251801} (\bibinfo {year} {2017})}\BibitemShut {NoStop}%
\bibitem [{\citenamefont {Adey}\ \emph {et~al.}(2019)\citenamefont {Adey} \emph
  {et~al.}}]{DB_2019}%
  \BibitemOpen
  \bibfield  {author} {\bibinfo {author} {\bibfnamefont {D.}~\bibnamefont
  {Adey}} \emph {et~al.} (\bibinfo {collaboration} {Daya Bay Collaboration}),\
  }\bibfield  {title} {\bibinfo {title} {Extraction of the $^{235}\mathrm{U}$
  and $^{239}\mathrm{Pu}$ antineutrino spectra at {D}aya {B}ay},\ }\href
  {https://doi.org/10.1103/PhysRevLett.123.111801} {\bibfield  {journal}
  {\bibinfo  {journal} {Phys. Rev. Lett.}\ }\textbf {\bibinfo {volume} {123}},\
  \bibinfo {pages} {111801} (\bibinfo {year} {2019})}\BibitemShut {NoStop}%
\bibitem [{\citenamefont {Bak}\ \emph {et~al.}(2019)\citenamefont {Bak} \emph
  {et~al.}}]{RENO_2019}%
  \BibitemOpen
  \bibfield  {author} {\bibinfo {author} {\bibfnamefont {G.}~\bibnamefont
  {Bak}} \emph {et~al.} (\bibinfo {collaboration} {RENO Collaboration}),\
  }\bibfield  {title} {\bibinfo {title} {Fuel-composition dependent reactor
  antineutrino yield at {RENO}},\ }\href
  {https://doi.org/10.1103/PhysRevLett.122.232501} {\bibfield  {journal}
  {\bibinfo  {journal} {Phys. Rev. Lett.}\ }\textbf {\bibinfo {volume} {122}},\
  \bibinfo {pages} {232501} (\bibinfo {year} {2019})}\BibitemShut {NoStop}%
\bibitem [{\citenamefont {Almaz\'an}\ \emph {et~al.}(2020)\citenamefont
  {Almaz\'an} \emph {et~al.}}]{STEREO_2020}%
  \BibitemOpen
  \bibfield  {author} {\bibinfo {author} {\bibfnamefont {H.}~\bibnamefont
  {Almaz\'an}} \emph {et~al.} (\bibinfo {collaboration} {STEREO
  Collaboration}),\ }\bibfield  {title} {\bibinfo {title} {Accurate measurement
  of the electron antineutrino yield of $^{235}\mathrm{U}$ fissions from the
  {STEREO} experiment with 119 days of reactor-on data},\ }\href
  {https://doi.org/10.1103/PhysRevLett.125.201801} {\bibfield  {journal}
  {\bibinfo  {journal} {Phys. Rev. Lett.}\ }\textbf {\bibinfo {volume} {125}},\
  \bibinfo {pages} {201801} (\bibinfo {year} {2020})}\BibitemShut {NoStop}%
\bibitem [{\citenamefont {Onillon}\ and\ \citenamefont
  {Letourneau}(2019)}]{Onillon+_2019}%
  \BibitemOpen
  \bibfield  {author} {\bibinfo {author} {\bibfnamefont {A.}~\bibnamefont
  {Onillon}}\ and\ \bibinfo {author} {\bibfnamefont {A.}~\bibnamefont
  {Letourneau}},\ }\bibfield  {title} {\bibinfo {title} {Investigation of the
  \uppercase{ILL} spectra normalization},\ }in\ \href@noop {} {\emph {\bibinfo
  {booktitle} {Applied Antineutrino Physics 2018 Proceedings}}}\ (\bibinfo
  {year} {2019})\ \Eprint {https://arxiv.org/abs/1911.06834} {arXiv:1911.06834
  [hep-ex]} \BibitemShut {NoStop}%
\bibitem [{\citenamefont {Hayes}\ \emph {et~al.}(2018)\citenamefont {Hayes},
  \citenamefont {Jungman}, \citenamefont {McCutchan}, \citenamefont {Sonzogni},
  \citenamefont {Garvey},\ and\ \citenamefont {Wang}}]{Hayes+_2018}%
  \BibitemOpen
  \bibfield  {author} {\bibinfo {author} {\bibfnamefont {A.~C.}\ \bibnamefont
  {Hayes}}, \bibinfo {author} {\bibfnamefont {G.}~\bibnamefont {Jungman}},
  \bibinfo {author} {\bibfnamefont {E.~A.}\ \bibnamefont {McCutchan}}, \bibinfo
  {author} {\bibfnamefont {A.~A.}\ \bibnamefont {Sonzogni}}, \bibinfo {author}
  {\bibfnamefont {G.~T.}\ \bibnamefont {Garvey}},\ and\ \bibinfo {author}
  {\bibfnamefont {X.~B.}\ \bibnamefont {Wang}},\ }\bibfield  {title} {\bibinfo
  {title} {Analysis of the {D}aya {B}ay reactor antineutrino flux changes with
  fuel burnup},\ }\href {https://doi.org/10.1103/PhysRevLett.120.022503}
  {\bibfield  {journal} {\bibinfo  {journal} {Phys. Rev. Lett.}\ }\textbf
  {\bibinfo {volume} {120}},\ \bibinfo {pages} {022503} (\bibinfo {year}
  {2018})}\BibitemShut {NoStop}%
\bibitem [{\citenamefont {{Kopeikin}}(2012)}]{Kopeikin_2012}%
  \BibitemOpen
  \bibfield  {author} {\bibinfo {author} {\bibfnamefont {V.~I.}\ \bibnamefont
  {{Kopeikin}}},\ }\bibfield  {title} {\bibinfo {title} {{Flux and spectrum of
  reactor antineutrinos}},\ }\href {https://doi.org/10.1134/S1063778812020123}
  {\bibfield  {journal} {\bibinfo  {journal} {Physics of Atomic Nuclei}\
  }\textbf {\bibinfo {volume} {75}},\ \bibinfo {pages} {143} (\bibinfo {year}
  {2012})}\BibitemShut {NoStop}%
\bibitem [{\citenamefont {Fayans}(1985)}]{Fayans_1985}%
  \BibitemOpen
  \bibfield  {author} {\bibinfo {author} {\bibfnamefont {S.~A.}\ \bibnamefont
  {Fayans}},\ }\bibfield  {title} {\bibinfo {title} {Radiative corrections and
  recoil effects in the reaction $\bar{\nu}_e + p \to n + e^+$ at low
  energies},\ }\href@noop {} {\bibfield  {journal} {\bibinfo  {journal} {Sov.
  J. Nucl. Phys.}\ }\textbf {\bibinfo {volume} {42}},\ \bibinfo {pages} {590}
  (\bibinfo {year} {1985})}\BibitemShut {NoStop}%
\bibitem [{\citenamefont {Vogel}(1984)}]{Vogel_1984}%
  \BibitemOpen
  \bibfield  {author} {\bibinfo {author} {\bibfnamefont {P.}~\bibnamefont
  {Vogel}},\ }\bibfield  {title} {\bibinfo {title} {Analysis of the
  antineutrino capture on protons},\ }\href
  {https://doi.org/10.1103/PhysRevD.29.1918} {\bibfield  {journal} {\bibinfo
  {journal} {Phys. Rev. D}\ }\textbf {\bibinfo {volume} {29}},\ \bibinfo
  {pages} {1918} (\bibinfo {year} {1984})}\BibitemShut {NoStop}%
\bibitem [{\citenamefont {{Kopeikin}}\ and\ \citenamefont
  {{Skorokhvatov}}(2017)}]{Kopeikin+_2017}%
  \BibitemOpen
  \bibfield  {author} {\bibinfo {author} {\bibfnamefont {V.~I.}\ \bibnamefont
  {{Kopeikin}}}\ and\ \bibinfo {author} {\bibfnamefont {M.~D.}\ \bibnamefont
  {{Skorokhvatov}}},\ }\bibfield  {title} {\bibinfo {title} {{Special features
  of the inverse-beta-decay reaction proceeding on a proton in a
  reactor-antineutrino flux}},\ }\href
  {https://doi.org/10.1134/S1063778817020223} {\bibfield  {journal} {\bibinfo
  {journal} {Physics of Atomic Nuclei}\ }\textbf {\bibinfo {volume} {80}},\
  \bibinfo {pages} {266} (\bibinfo {year} {2017})}\BibitemShut {NoStop}%
\end{thebibliography}%

\end{document}